\documentstyle[epsfig,longtable]{aipproc}

\begin{document}
\title{ Observation of Strong Variability in the X-Ray
                      Emission from Markarian 421 Correlated with the
                      May 1996 TeV Flare}
\author{Michael Schubnell\thanks{Support for
this work was provided in part by NASA contract NAG 5-3264}}
\address{Randall Laboratory of Physics\\
University of Michigan, Ann Arbor, Michigan 48109}

\maketitle

\begin{abstract}
We observed the BL Lac object Markarian 421 with the X-ray
satellite RXTE and the Whipple Air Cerenkov Telescope during
a two week correlated X-ray/gamma-ray campaign in May 1996.
Two dramatic outbursts with extremely rapid and strong flux variations
were observed at TeV energies during this period.
The X-ray emission in the 2-10 keV band was highly variable
and reached a peak flux of $5.6\times10^{-10}$ erg cm$^{-2}$ s$^{-1}$,
a historic high. Similar behavior was observed for the TeV
emission. In contrast to
earlier near-simultaneous X-ray/gamma-ray observations of Mrk 421, the
variability amplitude is much larger at TeV than at X-ray energies. This
behavior is expected in Synchrotron Self-Compton models.
\end{abstract}

\section*{Introduction}

Present blazar research focuses on the question of
how relativistic jets are formed and how particles are accelerated
to energies beyond a TeV.
To find an answer to this question we are trying to understand the very
basic properties of the jet, i.e. particle population, velocities,
total energy, and magnetic field strength.
The non-thermal blazar emission from radio to UV is generally thought to be
synchrotron radiation beamed from a relativistic jet viewed at small angles.
In the case of Markarian 421, the synchrotron emission dominates up to
X-ray energies. A second component, observed above the synchrotron break,
is typically relatively flat and extends to X-ray and sometimes
to gamma-ray energies, as in the case for Markarian 421.

In the context of current theoretical models invoked to explain the
broadband energy spectrum of Markarian 421 and other similar X-ray
selected BL Lac objects, correlated observations can be used
to understand the physics of the electron population believed to be the
progenitor of the high energy emission.
The energy spectrum near the spectral break around a few keV may
give a handle on the maximum electron energy in the system.
Mufson et al. (1990) report a two-component X-ray spectrum with a steepening
tail above 4.5 keV. This suggests the contribution of two different components
to the X-ray emission: synchrotron radiation from the highest energy
electrons and inverse-Compton scattered photons from the lowest energy
electrons. In such a scenario, the X-ray band plays a crucial role in the
study of the broadband emission \cite{urry96}.

Markarian 421 has been monitored previously and has shown rapid variability
on short time scales. Detailed studies based on EXOSAT observations showed
significant variation in the 0.5-10 keV band on time scales from several
hours to several days \cite{george88}. This strongly
suggests that typical X-ray high states are relatively short lived.
Near coincident flaring in the X-ray and TeV gamma-ray emission was observed
in 1994 and in 1995 \cite{buckley96,macomb95} which prompted us to
propose extensive observations with the newly launched X-ray satellite, XTE,
and the Whipple telescope.

\section*{Observations}

The X-ray observations discussed here were carried out with the PCA
(Proportional Counter Array) onboard the Rossi X-Ray Timing Explorer
\cite{bradt93}.
Between 1996 May 4 and May 21, the PCA was pointed on average 4 times
every day at Markarian 421. A typical observation resulted in a net
exposure of 600 seconds.

The individual X-ray spectra were fitted with a
single power-law model with energy spectral index $\alpha$,
$F(E)=F_0E^{-\alpha}$, with
the absorbing column $N_H$ fixed at the
Galactic value ($1.5\times10^{20} cm^{-2}$) \cite{elvis89}.
We found
that the simple power law function describes the data well in the energy range 
between 2 and 10 keV if we allow for a 2\% systematic error for energies
between 3.5 and 6 keV. This additional systematic error accounts for
small uncertainties
in the channel-to-energy conversion which was not finalized at the time of
this analysis.
Alternately, we also have applied a single
power-law model and allowed for a varying column density. While this
resulted in slightly improved $\chi^2$ values
for fits between 2 keV and 10 keV,
it also resulted in $N_H$ values $\approx$ 20 times larger than
the Galactic value.
A broken power-law model, successfully applied to ASCA
data \cite{takahashi96}, did not fit any of the spectra well.

\section*{Results}

During the observation period two very remarkable outbursts were
observed at TeV energies (May 7 and May 15). The May 7 flare
is the most intense flare recorded at TeV energies \cite{gaidos96}.
During the course of 2 hours of observing, the
gamma-ray emission increased steadily from 4 times to about 30 times the
average flux. The second TeV flare (May 15) was less pronounced but
shows an even shorter doubling time of $\approx$ 20 minutes.

\begin{figure}[h!] 
\centerline{\epsfig{file=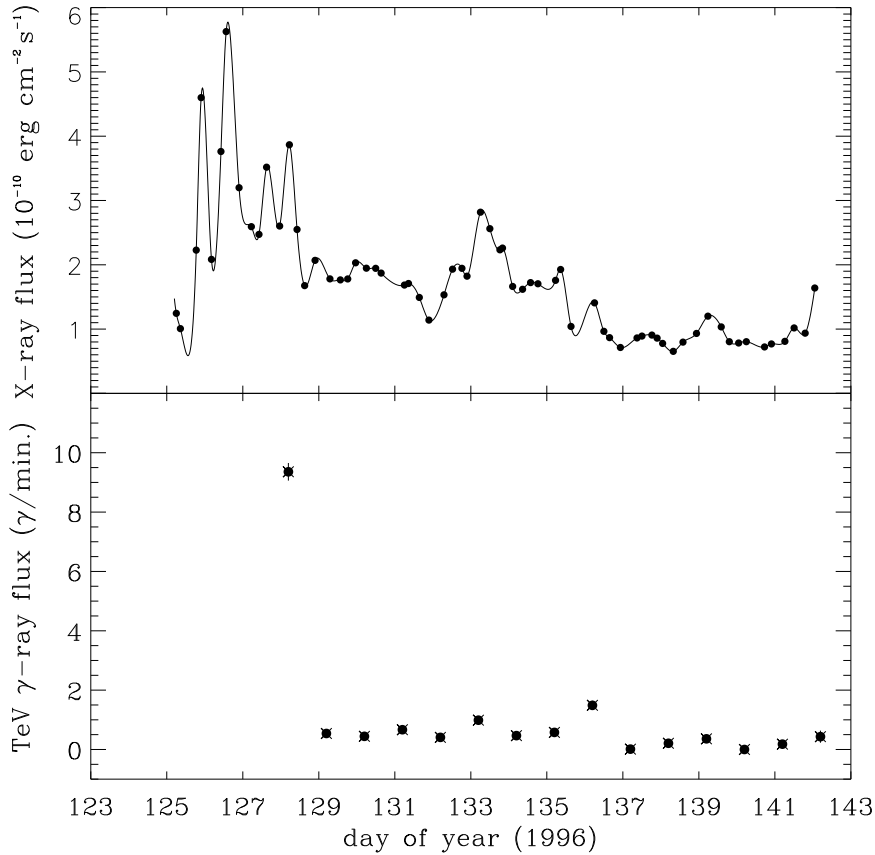,width=4.5in}}
\vspace{10pt}
\caption{The X-ray (2-10 keV) and TeV ($>$ 300 GeV) emission from
Markarian 421 during May 1996. We included a fitted cubic spline function
as a guidance for the eye.}
\label{fig1}
\end{figure}

The time history of the May 1996 X-ray and TeV observations is
shown in figure \ref{fig1}.
The X-ray light curve during the high state between
day 125 and day 129 shows very rapid quasi-periodic variations with
similar rise and fall times of the order
of several hours. This behavior is an indication
that geometrical effects may need to be considered in order to
explain the observed variability.
The 2-10 keV flux was high throughout the observing period
compared to previous measurements (see \cite{takahashi95} for a
compilation of the long term variability of Markarian 421). 
On May 5 (day 126) the 2-10 keV flux increased
to $5.6\times10^{-10}~erg~cm^{-2}s^{-1}$, brighter than all previous
observations. 

The TeV high state on May 7 (day 128) clearly occurred during strong
activity in the X-ray band.
Restricted by the bright moon, TeV observations could not be obtained
prior to day 128. This leaves some ambiguity in determining the exact
time delay between TeV and X-ray high state. However, the close coincidence
of the TeV and X-ray flare suggests that both arise from the same electron
population in the jet.

An interesting conclusion can be drawn from comparing the relative
amplitudes of the flare states in both bands. While previous flare
observations \cite{buckley96,macomb95} claim a comparable amplitude in the
variability, during this observation, the flux at TeV energies rose
a factor of $\approx$ 30 above the quiescent level, compared to a factor
of $\approx$ 5 for the keV flux. This is expected in Synchrotron
Self-Compton (SSC) models where the X-rays are the seed photons that
are Comptonized to produce the TeV flux. Simultaneous optical data taken
with the 48 inch optical telescope on Mt. Hopkins did not show an increase
in the flux during this period \cite{mcenry96}. This also confirms previous
observations where the flux at longer wavelengths remained
relatively constant during strong X-ray/TeV flares.

\begin{figure}[h!] 
\centerline{\epsfig{file=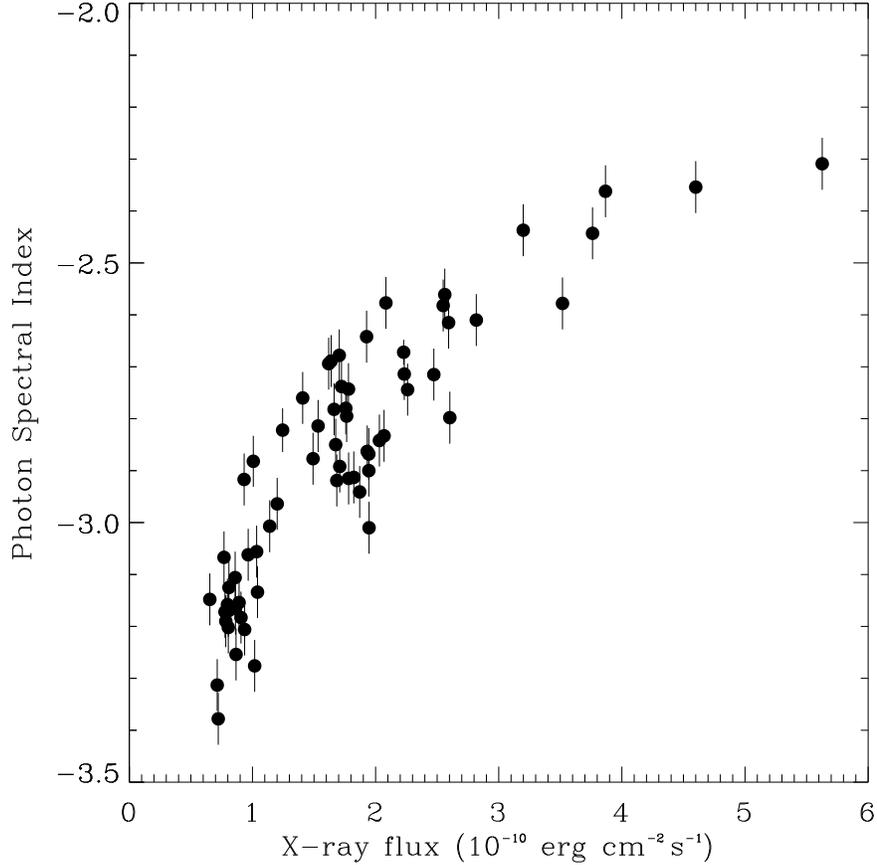,width=4.5in}}
\vspace{10pt}
\caption{
Correlation between the observed 2-10 keV flux and
the photon spectral index  $\alpha$ (A single power-law function with
index $\alpha$, such that the flux $\Phi$(E) = $\Phi_0$E$^{-\alpha}$,
describes the spectra well).}
\label{fig2}
\end{figure}

The variation in the observed flux shows a strong
correlation with the photon spectral index (figure \ref{fig2}). A spectral
hardening is observed during a phase of increased emission. This can be
explained by the injection of an electron component
more rapid than typical synchrotron cooling time scales \cite{tashiro95}.

\section*{Summary}

We detected strong variability in the 2-10 keV emission from
Markarian 421 during observations carried out in May 1996, parallel
to the TeV observations by the Whipple Collaboration. The
correlated variability of X-ray and
TeV emission
strongly supports models which involve synchrotron radiation and
the Inverse Compton (IC) process to describe the
spectral energy distributions of blazars. In the case where the
seed photons for the IC process are the synchrotron photons (SSC models),
the IC flare amplitude is expected to be proportional to the square of
the synchrotron flare amplitude. The presented data suggest this scenario,
in which, at
least in the case of Markarian 421, the variability in the TeV
luminosity (IC) is caused by variability of the X-ray photons (synchrotron).
The quasi-periodicity in the X-ray emission, with similar
time scales in the rise and fall times of the individual flares
indicate that geometric effects may have to be considered to understand
the perturbations which lead to the observed variability.

\acknowledgments

The author thanks the Whipple Collaboration for cooperation
concerning the TeV data, and J. Lochner, G. Rohrbach, and A. Rots of the
RXTE Guest Observer Facility for providing excellent technical support.
I also thank C. Akerlof, R. Sambruna, and M. Urry for helpful
discussions.


\begin{references}
\bibitem{buckley96} Buckley, J. H., et al., {\it ApJ} {\bf 472}, L9 (1996).
\bibitem{bradt93} Bradt, H. V.,
Rothschild, R. E., \& Swank, J. H., {\it A\&A} {\bf 97}, 355 (1993).
\bibitem{elvis89} Elvis, M.,
Lockman, F. J., \& Wilkes, B. J., {\it AJ} {\bf 97}, 777 (1989).
\bibitem{gaidos96} Gaidos, J. A., et al., {\it Nature} {\bf 383}, 319 (1996).
\bibitem{george88} George, U. M.,
Warwick, R. S., \& Bromage, G. E. {\it MNRAS} {\bf 232}, 793 (1988).
\bibitem{macomb95} Macomb, D. J., et al., {\it ApJ} {\bf 449}, L99 (1995).
\bibitem{mcenry96} McEnery, J., et al., to appear in proc. 25th ICRC (1997).
\bibitem{mufson90} Mufson, S. L., et al., {\it ApJ} {\bf 354}, 116 (1990).
\bibitem{takahashi95} Takahashi, T., et al., Mem. Soc. Astron. Ital.,
67, 533 (1996).
\bibitem{takahashi96} Takahashi, T., et al., {\it ApJ} {\bf 470}, L89 (1996).
\bibitem{tashiro95} Tashiro, M., et al., {\it PASJ} {\bf 47}, 131 (1995).
\bibitem{urry96} Urry, C. M., et al., {\it STSI preprint} 1017 (1996).
\end{references}
\end{document}